\documentclass{article}
\usepackage{amsmath}
\usepackage{graphicx} 
\usepackage{color}

\title{A minimal model of boosting and waning \\in a recurrent seasonal epidemic}

\date{}

\begin{document}

\maketitle
           
\Large\noindent Siyu Chen \\

\noindent\normalsize High Meadows Environmental Institute\\
Princeton University, Princeton, NJ 08544\\
{\tt{siyu.chen@princeton.edu}}

\  

and

\

\Large\noindent David Sankoff\\

\noindent\normalsize Department of Mathematics and Statistics\\
University of Ottawa\\
150 Louis Pasteur Pvt\\
Ottawa, Ontario, Canada K1N 6N5\\
{\tt{sankoff@uottawa.ca}}

\newpage

\begin{abstract}
We propose a model of the immunity to a cyclical epidemic disease taking account not only of seasonal boosts during the infectious season, but also of residual immunity remaining from one season to the next.  The focus is on the exponential waning process over successive cycles, imposed on the temporal distribution of infections or exposures over a season.  This distribution, interacting with the waning function, is all that is necessary to reproduce, in mathematically closed form, the mechanical cycle of boosting and waning immunity characteristic of recurrent seasonal infectious disease. Distinct from epidemiological models predicting numbers of individuals moving between infectivity compartments, our result enables us to directly estimate parameters of waning and the infectivity distribution. We can naturally iterate the cyclical process to simulate immunity trajectories over many years and thus to quantify the strong relationship between residual immunity and the time elapsed between annual infectivity peaks.

\end{abstract}

\section{Introduction}In the study of the residual immunity remaining after each cycle, often annual,  of a recurrent infectious disease, exemplified by influenza \cite{brauer}, the rate of resistance waning warrants  closer modeling than the customary linear decline over a few months or a year  (as in \cite{punyo}).  In this note we investigate an exponential decay of immunity, expressed in terms of abstract units of antibody titre, with parameter $\omega$.  In addition to this parameter, the analysis requires a functional form $f$ (e.g., Beta, truncated normal or gamma, modified cosine, etc.) with peak, mean or median $\mu$ falling somewhere during the infectivity season, multiplied by an amplitude $A_I$.

The focus is on the waning process over several cycles, taking into account as well the functional form, namely the temporal distribution, of infections over an infectious season and the severity of the epidemic.

The exponential decay of immunity after vaccination or during an epidemic season  has been studied in some detail, often in terms of proportions of the population moving between infectivity compartments \cite{antia, nz,elkhalifi}.  Although studies such as \cite{zhao} have empirically derived mathematical forms for the residual immunity from one season to the next, these do not model the process of waning, which is the topic of this paper.

Our approach is to formulate a model that contains a minimum of elements, namely a probability distribution representing the time course of the infectious period, including at least parameters for location, dispersion and amplitude as well as a waning parameter.   This contrasts with other models of the progress of an epidemic, involving multiple population compartments and many parameters interacting with the waning process.  With the minimal model, purely analytic closed-form solutions separating the effects of individual parameters can be derived. 

In the ensuing sections, we present the model, analyze the effects of the various parameters and suggest a maximum likelihood approach to the waning parameter. We then iterate many annual cycles of the process, based on a single random choice of input titre.  This leads to an explicit discovery of a linear quantitative relation between seasonal titre and the length of the interval between annual infectivity peaks.

\section{The model}The change of antibody titre between ``sampling times" $T_{I-1}$ and $T_I$ is represented by the 
equation: 
\begin{equation}\label{master}
    X_{T_I}=X_{T_{I-1}}e^{-\omega(T_I-T_{I-1})}+ \int_{T_{I-1}}^{T_I} A_If_\mu(t)e^{-\omega(T_I-t)}dt,
\end{equation}

where \begin{itemize} 
\item $T_{I-1}$ is end of the previous cycle, and the beginning of the current cycle, 
in months, e.g. $T_{0}=3$, meaning March 31,
\item $X_{T_{I-1}}$ is the titre at end of previous cycle, 
\item $T_I$ is end of the current cycle, in months, where $T_I=T_{I-1}+12$,
\item $X_{T_{I}}$ is the titre at end of the current cycle
\item $\mu$ or $\mu_I$ is the peak of the current infectious season (default $\mu_I=T_{I-1}+9$),
\item $\omega$ is the decay parameter (in months$^{-1}$) for  exponential waning (default $\omega=\frac{1}{24}$,
\item $A_I$ is the amplitude (default $= 1$) of the infectious season, 
\item $f_\mu(t)$ is a probability density, for example one based on a transformed cosine, with support between the infectious season onset date $t_o=\mu-\frac{\pi}{a}$ and end date $t_e=\mu+\frac{\pi}{a}$: 
\begin{equation}f_{\mu}=\frac{a}{2\pi}(1+ \cos a(t-\mu))\end{equation} for $t\in [\mu-\frac{\pi}{a},\mu+\frac{\pi}{a}]$, and $f_{\mu}=0$ elsewhere, where $a^{-1}$ is a measure of dispersion, in months.
\end{itemize}

Equation (\ref{master}) may be rewritten

\begin{equation}\label{reduced}
    X_{T_I}=X_{T_{I-1}}e^{-\omega(T_I-T_{I-1})}+ A_I{e^{-\omega T_I}}\int_{t_o}^{t_e} f_\mu(t)e^{\omega t}dt.
    \end{equation}
    The integral in this formula is
\begin{equation}\label{integrated}
    \int_{t_o}^{t_e} f_\mu(t)e^{\omega t}dt=
e^{\mu w}  \frac{\sinh(\pi\theta)}{\pi\theta( 1 +\theta^2)},
    \end{equation}
    where $\theta=\frac{\omega}{a}$, so that equation (\ref{master}) becomes
\begin{equation}\label{final}
    X_{T_I}=X_{T_{I-1}}e^{-\omega(T_I-T_{I-1})}+ A_I{e^{-\omega (T_I-\mu)}}  \frac{\sinh(\pi\theta)}{\pi\theta( 1 +\theta^2)}.
    \end{equation}

\section{The effect of the parameters}  To understand the role of each of the parameters in equation (\ref{master}), we first assign them default values, and then vary each one systematically to see how they affect the titre at $T_I$.

The defaults:
\begin{enumerate}
\item amplitude $A$:, default$=1$,
\item waning rate $\omega$: default $= \frac{1}{24}$ months$^{-1}$,
\item $f$ (cosine example parameters), season mean $\mu$: default$={T_{I-1}}+9$, diversity $a^{-1}=\frac{2}{\pi}$: in months. Note that $a$ and $\mu$ determine onset date $t_o=8$ and end date $  t_e=12$, since $a=\pi/(\mu-t_o)=\pi/(t_e-\mu)$,
\item $T_{I-1},T_I$, (annual sampling dates, in months: 15, 27, 39,$\dots$
\end{enumerate}

\subsection{Domain of applicability of the model} With the default parameters specified above, we require $t_e \le t_o + 12$ to ensure that infectious seasons do not overlap. 
\begin{figure}[h]\label{regress}
\centering
\vspace{0.2in}
\includegraphics[width=0.48\textwidth]{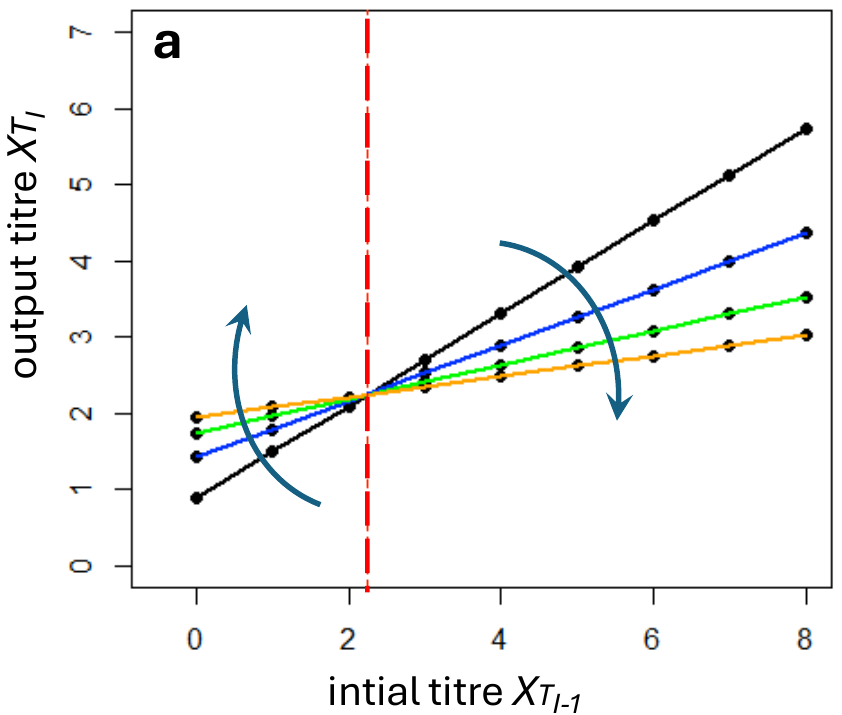}
\includegraphics[width=0.47\textwidth]{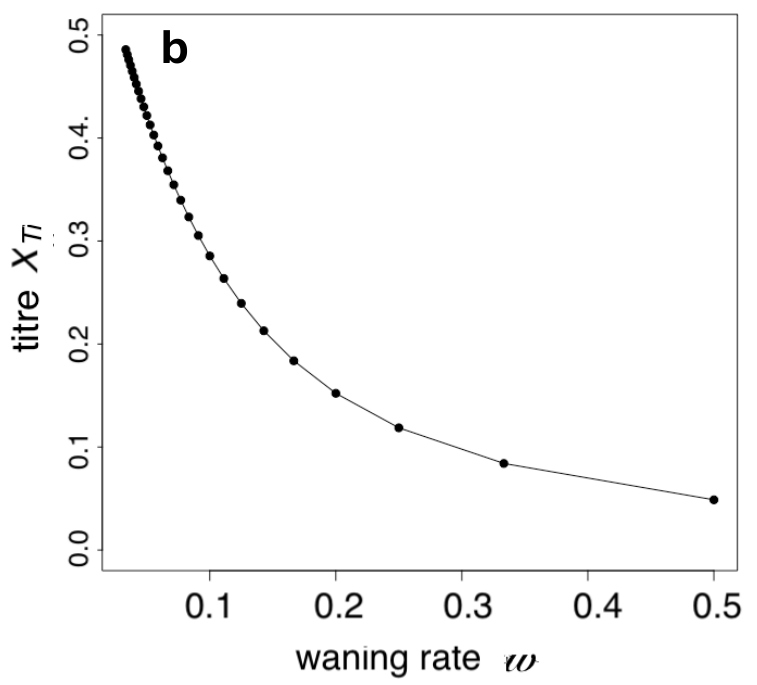} 
\includegraphics[width=0.46\textwidth]{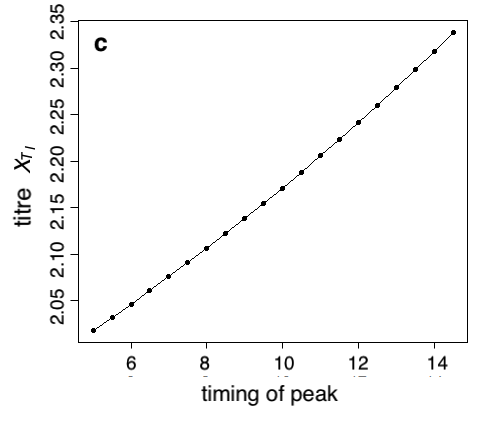} \ \ \ 
\includegraphics[width=0.45\textwidth]{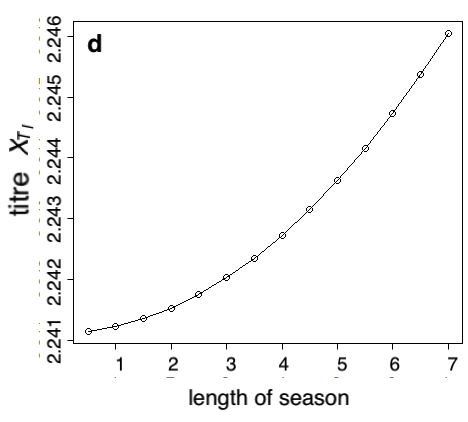}\caption{Effects of parameters: (a) The dots stacked above each initial titre represent successive iterations of the default model. (b) Effect of the waning parameter $\omega$ on $X_{T_I}$. (c) Effects of early or delayed season.  (d) Length of season and waning}\label{regress}
\end{figure}
Whatever the ``input" $X_{T_{I-1}}$ into the default model, the ``output" regresses to an equilibrium value of 2.24$\dots$ as depicted in Figure \ref{regress}a. Since titre is a linear function of amplitude, this regression behavior is preserved even when the amplitude $A$ changes, although with a different equilibrium value.  There is no upper bound to $A$ or to $X_{T_{I-1}}$ that restricts the calculation of equation (\ref{final}).

\subsection{The effect of waning} The role of waning in limiting residual immunity in a seasonally recurring infectious disease is the central motivation for this paper. Figure \ref{regress}b shows a decidedly non-linear effect of the parameter It can be see that for very high waning rates, the output titre is sharply reduced, but lower values of $\omega$ result in $X_{T_I}$ conserving much of the input titre and even surpass it.

The model parameters $\mu$, $t_o$, $t_e$ and $A$ are relatively accessible to direct measurement through public health statistics \cite{can,usa}, but $\omega$ can only be inferred indirectly. Its importance to the modeling of residual immunity seen in Figure \ref{regress}b raises the questions of how to estimate it. This will be the topic of Section \ref{estimation} below.
\subsection{Seasonality effects} An early or late season will increase or decrease the time until the $T_I$ sampling, increasing or decreasing waning time, respectively, as seen in Figure \ref{regress}c.  This effect manifests a largely linear response, and is substantial.  

Another aspect of the seasonality function $f$ is the concentration versus the extension of the season.  In the default model, this is embodied in the parameter $a$, which determines whether the season will extend over just a few weeks, or will spread over months longer than usual.

Hampering the analysis of the seasonality effects is the requirement  that $t_e\le T_I$, otherwise the season would last beyond the sampling date and the integration of $f$ would extend into future seasons. This limits our ability to investigate the effects of $\mu$.  To accommodate this problem within our model, one solution would be to compress the season by increasing the parameter $a$ so that a larger range of values of $\mu$ can be considered.  Thus for a given $\mu$, if $t_e>T_I$ we redefine a new value
\begin{equation}
a^*=\frac{\pi}{T_I-\mu},
\end{equation}
so that the interval of integration is $[\mu-\frac{\pi}{a^*}, \mu+\frac{\pi}{a*}]$.  

A similar adjustment could be defined for a season starting early, where
\begin{equation}
a^{**}=\frac{\pi}{\mu-T_{I-1}},
\end{equation}
so that the interval of integration is $[\mu-\frac{\pi}{a^{**}}, \mu+\frac{\pi}{a^{**}}]$.  

An alternate way of allowing later (or early) seasons without artificially tampering with the shape of the distribution would be to allow sampling during the season.  This solution, however, would involve the high in-season titres, which would mask the inter-season waning process we are studying.  

Another solution is to adapt the sampling time to the end of the season. We explore this in Section \ref{multiyear} below.  In any case, the effect of season length, for a fixed $\mu$ and a fixed total infections or exposure, is miniscule, as in Figure \ref{regress}d.

\section{Estimating $\omega$}\label{estimation}
\begin{figure}[h!]
\vspace{-1in}
\centering \hspace{-.2in}
\includegraphics[width=.5\textwidth]{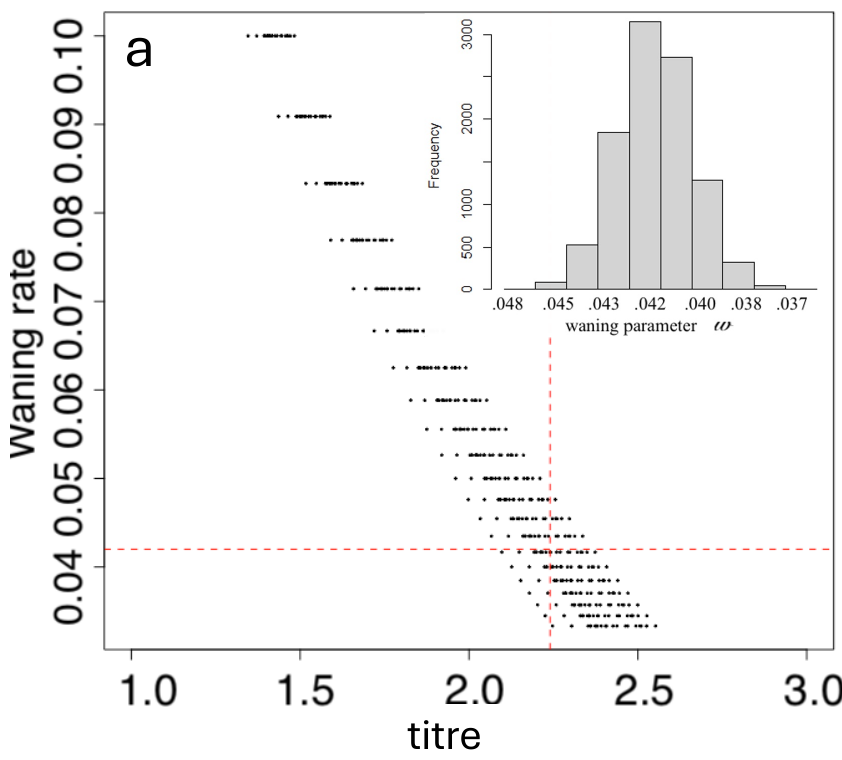}
\vspace{.2in}
\includegraphics[width=0.50\textwidth]{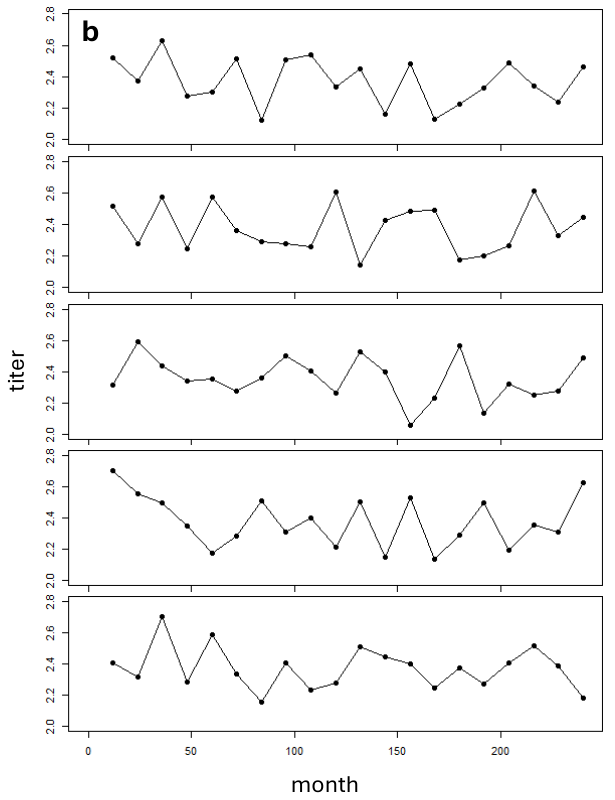}
\includegraphics[width=0.65\textwidth]{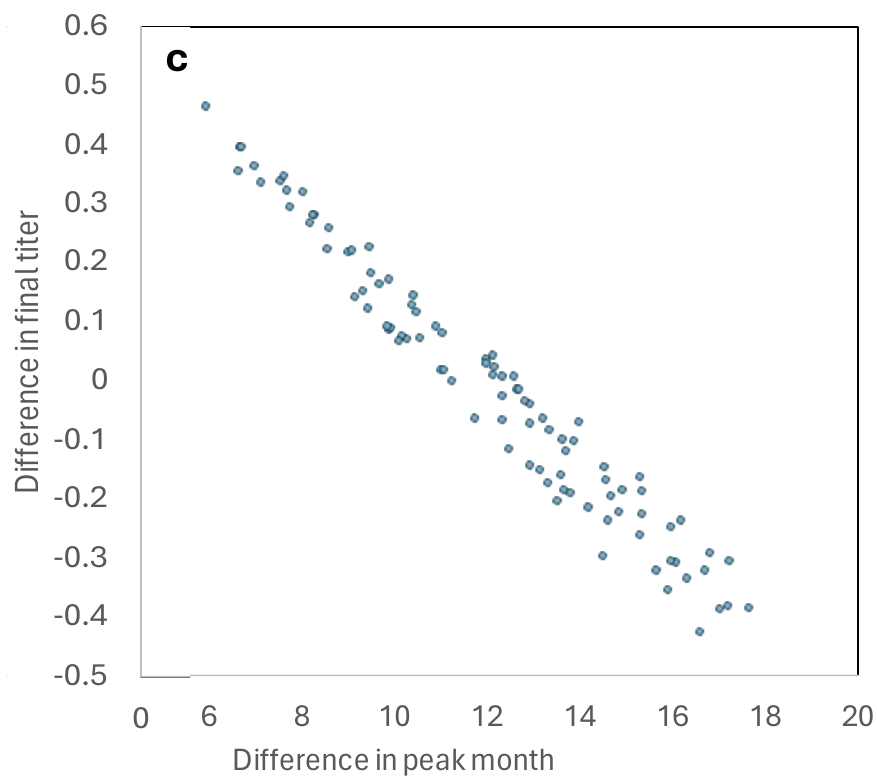}\caption{(a) Waning parameter by output titre. Input contains normal error $\sigma=0.8$ Vertical line above $T_I=2.24$ show the distribution of estimates of $\omega$. Inset: Likelihood of  $\omega$. (b) Sample trajectories of 20 seasons produced by iterating the model. (c) Association of titre change with length of inter-season times.}\label{multi}
\end{figure}

The crucial question of estimating $\omega$ may be addressed by simulating $T_2$, as in Figure \ref{regress}b.  Interchanging axes of this graph, as in Figure \ref{multi}a, allows us to visualize the range of values of $\omega$ associated with each end-of-season titre $X_{T_2}$.

Another way of looking at estimation error with fixed $\omega$ and $\mu$ is to introduce measurement error on the $X_{T_2}$.  The results of this are depicted in Figure \ref{multi}a.  Here data summarized by the depicted distribution were generated using the default parameter values, with the addition of error in the observed output titres. To convert it into a likelihood function the value of $\omega$ at each point on the waning axis  replaces equilibrium titre values at that point.

\section{Multiyear trends}\label{multiyear} To further investigate the stability of our model, we carried out a simulation  of 20 recurrent infectious seasons interspersed with quiescent periods for the rest of each cycle (i.e., year). 

Initialized with a random titre at date $T_1$ corresponding to the end of a typical infectious period, the peak infection date $m$ was chosen from a uniform probability over a wide range, September to April, and the first iteration was performed with $T_2$ set to be $m+\frac{\pi}{a}$, with output $X_{T_2}$.

The output $X_{T_2}$ at time $T_2$ from the first cycle, is then used as the starting titre $X_{T_3}$ at time $T_3$ of the second cycle. The peak of infections is again randomly chosen from September to April.

This calculation is repeated for the third and subsequent cycles.
Figure \ref{multi}b shows five typical trajectories of $X_{T}$ over the 20-year span.

We then plotted the change in titre $\Delta_X=X_{T_I}-X_{T_{I-1}}$ as a function of $\Delta_\mu=\mu_I-\mu_{I-1}$, the shift in peak infections between the $I-1$-st and $I$-th cycle.  The results in Figure \ref{multi}c show a tight linear relation between the two quantities.  The slope is -0.9 titre units/12 months differential, or 0.075 units/month.  This compares to $\log\omega = 0.042 $/month.

\section{Discussion and conclusions}
We have proposed a minimal model which takes into account titre increases during one season of an infectious disease, waning, and variable seasonality.  All of these are necessary to reproduce, in an elemental form, the mechanical cycle of boosting and waning characteristic of recurrent seasonal infectious disease -  and nothing else is necessary to evoke this pattern. 

The opposing tendencies due to boosting and waning is essentially due to the waning parameter $\omega$  and the location parameter $\mu$ of the infectious season.  

The dispersion parameter $a$ has little effect, and can vary considerably without materially affecting the change in titre from one cycle to the other.

The amplitude $A$, measuring the severity of the infectious season, is related linearly to the value of the equilibrium titre.  It does not interact mathematically with the other parameters of the model.

By relaxing the assumption of fixed sampling times, we could explore the variation in titre as a function of the peak location parameter $\mu$.  This revealed a strong association between shift of peak month and change of titre at the end of the season.

Although our discussion has been phrased in terms of immunity, sampling times and antibody titres, the  generality of our model actually means that it is not specifically attuned to any specific aspect of the infectious disease season or yearly cycle, such as exposures, infections,  symptoms, seroprevalence or antibody levels, as long as the annual boost can be represented by a distribution. Indeed, only the specific  peak season period  is critical; the shape of the distribution is less so.

Furthermore, we have not assumed anything about viral strains.  Apparent waning associated with the disease may reflect mutational drift or selection in the antigen, rather than immunological processes per se.  This does not distract from the pertinence of our model is describing the periodic behaviour of boosting and waning.

\end{document}